\documentclass[aps,altaffillsymbol,prl,twocolumn,superscriptaddress,showpacs,nofootinbib]{revtex4-1}

\usepackage{graphicx}	
\usepackage{dcolumn}	
\usepackage{bm}			
\usepackage[export]{adjustbox}
\usepackage{units}
\usepackage{float}
\usepackage{textcomp}
\usepackage{amssymb}
\usepackage{datetime}
\usepackage{mathtools}
\usepackage{multirow}
\usepackage{amsmath}
\usepackage{tabularx}
\usepackage{enumitem}
\usepackage{booktabs}
\usepackage{appendix}
\usepackage{color,soul}
\usepackage{lineno}

\begin{document}

\title{Magnetic moments of short-lived nuclei with part-per-million accuracy \\
\small Towards novel applications of $\beta$-detected NMR in physics, chemistry and biology}

\author{R. D. Harding}\altaffiliation{Current address: Nottingham University Hospitals Trust, Nottingham, United Kingdom}
\address{CERN, Geneva, Switzerland}
\address{University of York, York, United Kingdom}
\author{S. Pallada} \altaffiliation{Current address: HES-SO, Geneva, Switzerland}
\address{CERN, Geneva, Switzerland}
\author{J. Croese}
\address{CERN, Geneva, Switzerland}
\address{University of Geneva, Geneva, Switzerland}
\author{A. Antu{\v s}ek}
\address{Slovak University of Technology, Bratislava, Slovakia}
\author{M. Baranowski}
\address{Adam Mickiewicz University, Poznan, Poland}
\author{M. L. Bissell}
\address{University of Manchester, Manchester, United Kingdom}
\author{L. Cerato}
\address{University of Geneva, Geneva, Switzerland}
\author{K. M.  Dziubinska-K{\"u}hn}
\address{Leipzig University, Leipzig, Germany}
\address{CERN, Geneva, Switzerland}
\author{W. Gins}\altaffiliation{Current address: University of Jyv{\"a}skyl{\"a}, Jyv{\"a}skyl{\"a}, Finland}
\address{KU Leuven, Leuven, Belgium}
\author{F. P. Gustafsson}
\address{KU Leuven, Leuven, Belgium}
\author{A. Javaji} 
\altaffiliation{Current address: University of British Columbia, Vancouver, Canada}
\address{CERN, Geneva, Switzerland}
\address{Oldenburg University, Oldenburg, Germany}
\author{R. B. Jolivet}
\address{University of Geneva, Geneva, Switzerland}
\author{A. Kanellakopoulos}
\address{KU Leuven, Leuven, Belgium}
\author{B. Karg}
\address{University of Geneva, Geneva, Switzerland}
\author{M. Kempka}
\address{Adam Mickiewicz University, Poznan, Poland}
\author{V. Kocman}
\address{National Institute of Chemistry, Ljubljana, Slovenia}
\author{M. Kozak}
\address{Adam Mickiewicz University, Poznan, Poland}
\address{Jagiellonian University, Krakow, Poland}
\author{K. Kulesz}
\address{University of Geneva, Geneva, Switzerland}
\address{CERN, Geneva, Switzerland}
\author{M. Madurga Flores}
\address{University of Tennessee, Knoxville, USA}
\author{G. Neyens}
\address{KU Leuven, Leuven, Belgium}
\address{CERN, Geneva, Switzerland}
\author{R. Pietrzyk }
\address{Adam Mickiewicz University, Poznan, Poland}
\author{J. Plavec}
\address{National Institute of Chemistry, Ljubljana, Slovenia}
\author{M. Pomorski}\altaffiliation{Current address: Univ. Bordeaux, CNRS, CENBG, UMR 5797, F-33170 Gradignan, France} 
\address{University of Warsaw, Warsaw, Poland}
\author{A. Skrzypczak}
\address{Poznan University of Technology, Poznan, Poland}
\author{P. Wagenknecht}\altaffiliation{Current address: University of Tennessee, Knoxville, USA}
\address{CERN, Geneva, Switzerland}
\address{Oldenburg University, Oldenburg, Germany}
\author{F. Wienholtz}\altaffiliation{Current address: Technical University Darmstadt, Germany}
\address{CERN, Geneva, Switzerland} 
\author{J. Wolak}
\address{Adam Mickiewicz University, Poznan, Poland}
\author{Z. Xu}
\address{University of Tennessee, Knoxville, USA}
\author{D. Zakoucky}
\address{Czech Academy of Sciences, Rez, Czech Republic}
\author{M. Kowalska} \thanks{Corresponding author: kowalska@cern.ch}

\address{CERN, Geneva, Switzerland}
\address{University of Geneva, Geneva, Switzerland}

\date{\today}

\begin{abstract}
We determine for the first time the magnetic dipole moment of a short-lived nucleus with part-per-million (ppm) accuracy. To achieve this two orders of magnitude improvement over previous studies, we implement a number of innovations into our $\beta$-detected Nuclear Magnetic Resonance ($\beta$-NMR) setup at ISOLDE/CERN. 
Using liquid samples as hosts we obtain narrow, sub-kHz linewidth, resonances, while a simultaneous {\it in-situ} $^1$H NMR measurement allows us to calibrate and stabilize the magnetic field to ppm precision, thus eliminating the need for additional $\beta$-NMR reference measurements. Furthermore, we use {\it ab initio} calculations of NMR shielding constants to improve the accuracy of the reference magnetic moment, thus removing a large systematic error. We demonstrate the potential of this combined approach with the 1.1 s half-life radioactive nucleus $^{26}$Na, which is relevant for biochemical studies. Our technique can be readily extended to other isotopic chains, providing accurate magnetic moments for many short-lived nuclei. Furthermore, we discuss how our approach can open the path towards a wide range of applications of the ultra-sensitive $\beta$-NMR in physics, chemistry, and biology.
\end{abstract}


\maketitle

\section{Introduction}

The magnetic dipole moment $\mu$ is a fundamental property of atomic nuclei, and it is one of the primary observables used to investigate the nuclear wave-function \cite{Neyens2003NuclearNuclei,NeugartNeyens2006-moments,Neyens2005MeasurementState,Asahi2001-NPA-mu-bNMR-nucl_str,Ohtsubo2012-49Sn,Yang2016-79Zn,Thielking2018-229mTh,Minkov2019-229mTh-theory,Ichikawa2019,raeder2020-PRL-No}. At the same time, it serves as a versatile probe to measure the local magnetic field at the nucleus. This ability lies at the core of various spectroscopic techniques, among which a prominent role is played by nuclear magnetic resonance (NMR), which is an indispensable tool for determining structural details and dynamics in chemistry, biology and materials science  \cite{Webb2018-NMR-applic,Naito2018-NMR-applic}.

In NMR experiments, one measures the Larmor frequency $\nu_L$ of nuclei of spin $I$ precessing in a magnetic field. This frequency is the product of the gyromagnetic ratio $\gamma$ of the nucleus and the local magnetic field $B$ at the site of the nucleus, i.e. the applied magnetic field corrected for the effect of the electrons in the sample.
\begin{equation}
    \nu_L = \frac{\gamma B} {2 \pi} = \frac{\mu B}{h I},
    \label{Eqn:larmor}
\end{equation}
If one wants to employ NMR to extract a nuclear magnetic moment $\mu$, two inputs are thus essential. First, the Larmor frequency $\nu_L$ must be measured. For stable nuclei $\nu_L$ has been determined with sub-ppm precision since the early years of NMR \cite{Purcell1953,Baker1963-Tl,Epperlein1973-Zn,Lutz1978-ZPhysA-63-65Cu}, thus is not the dominant source of uncertainty in the derived magnetic moment.  
The second essential input to derive $\mu$ is the NMR shielding, describing the local effect of electrons in the sample on the applied magnetic field. Until recently, this effect has been poorly quantified and sometimes even neglected~\cite{Stone2005}. However, the introduction of reliable NMR shielding constants, provided by modern \textit{ab initio} methods~\cite{Helgaker1999,Malkinbook2004} enabled correction of this source of the systematic error in nuclear magnetic moment data~\cite{aakjmjwmmwcpl411,mjaapgkjwmmwpnmrs67}, which in extreme cases reached per-mill or per-cent levels~\cite{C6CP01781A,Skripnikov2018,Antusek2018,D0CP00115E}.
This correction turned out to be crucial for the tests of QED in the strong electromagnetic fields of highly-charged ions~\cite{mghgammppra58,Indelicato2019}.
Here, a more accurate value of the magnetic moment of $^{209}$Bi ~\cite{Skripnikov2018,Antusek2018} resolved a significant discrepancy between the measured and predicted hyperfine splitting (an effect of the interaction between $\mu$ and the magnetic field produced by the atomic electrons) of highly-charged $^{209}$Bi~\cite{Ullmann2017}, showing that QED is still valid in such a strong magnetic field.
Based on the corrected magnetic moments a new referencing scheme in NMR spectroscopy was also proposed~\cite{Jackowski2010b}, which allows a direct measurement of the NMR shielding instead of a chemical shift (i.e. a difference in NMR shieldings in different hosts).

More accurate nuclear magnetic moments can clearly bring new applications in different fields of research, as shown above for stable nuclei. However, magnetic moments of short-lived nuclei have not yet been measured with equally small uncertainty. In the present work, we demonstrate for the first time the determination of a magnetic moment of a short-lived nucleus with ppm accuracy. This has been achieved for the 1.1 s half life $^{26}$Na using an improved version of the $\beta$-NMR technique, combined with {\it ab initio} calculations of NMR shielding for the stable reference $^{23}$Na. The isotope $^{26}$Na was used for the proof-of-principle experiment, because of sodium importance for biochemistry applications \cite{Karg2020proposal}.

The $\beta$-NMR technique is based on the directional asymmetry of $\beta$-particle emission from spin-polarized $\beta$-decaying nuclei \cite{Bloembergen-betaNMR-53,shirley-betaNMRAnalChem69}. The most attractive feature of the method is its sensitivity, which is up to $10^{10}$ times higher than in conventional NMR \cite{Gottberg2014}, with down to $10^6$ resonating nuclei leading to an NMR spectrum. The technique has been applied to measure the magnetic moments of short-lived nuclei down to per-mill precision~\cite{connor-prl-8Li-moment-1959,WINNACKER1976-bNMR-neutr-mu-Ag,Arnold1987NUCLEARLlLi,OKUNO1995-14-15B-betaNMR-moment,Onishi1995-NPA-9C-mu-bNMR,Minamisono1996-9C13O-betaNMR,ueno1996-prc-17N-17B-moments-projectile,Geithner1999MeasurementNucleus11Be,Keim2000,ogawa2002-PrThPhys-17C-spin-moment,Neyens2005MeasurementState} and for structural investigations in materials science~\cite{MIHARA2007-12N-PhysB-betaNMR-material,Salman2012,MacFarlane2014,McKenzie2014,Cortie2015,MacFarlane2015,Cortie2016,PhysRevB.96.094402}. However, in chemistry and biology, $\beta$-NMR is far from being a routinely applicable spectroscopic method ~\cite{Gottberg2014,Sugihara2017NMRWater,Szunyogh2018}, due to numerous experimental challenges. One of them is the requirement of time-consuming reference measurements with the same short-lived nucleus in a different chemical environment~\cite{Sugihara2017NMRWater,MacFarlane_2014-8Li-chem-shift}. Furthermore, those reference measurements are performed in a solid-state sample ~\cite{Szunyogh2018,Sugihara2017NMRWater}, resulting in relatively wide resonance signals, thus increasing the final error on the extracted experimental value. Another challenge is due to the reduced precision and accuracy in the measured frequency and deduced magnetic moments, which prevent a direct comparison of the data with results from conventional NMR and from {\it ab initio} chemical calculations of local fields. The work presented here addresses all of the above limitations of the $\beta$-NMR technique, with the key ingredient being an accurate magnetic moment measurement.

The developments presented here will be crucial for future applications of high-precision $\beta$-NMR spectroscopy using a variety of radioactive probes, not only in the fields of chemistry and biology, but also for nuclear structure research. For example, determining the neutron distribution in light neutron-rich nuclei \cite{Puchalski2014,Takamine2016} is experimentally very challenging as neutrons do not carry electric charge. However, one can access this information by measuring the distribution of magnetization inside exotic nuclei. This requires very high-precision magnetic moment measurements, combined with high-precision hyperfine-structure measurements on the same isotope, to be sensitive to the `hyperfine anomaly' \cite{Bohr1950, Karpeshin2015,Persson2018}.

\section{Techniques}

\subsection{$\beta$-NMR on short-lived $^{26}$Na}
\begin{figure*}[t]
    \centering
        \includegraphics[width=\textwidth]{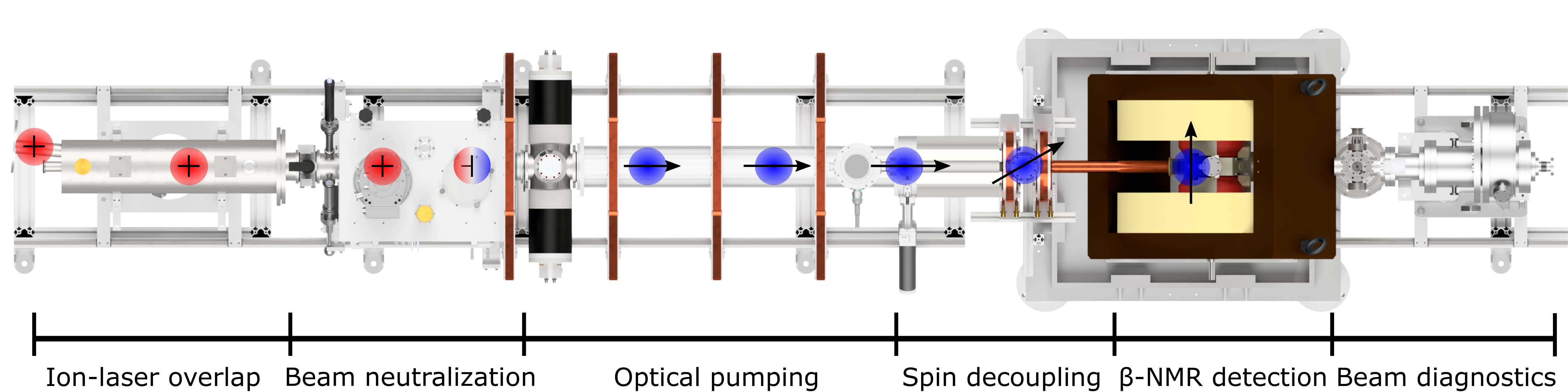}
        \caption{Top view of the laser-polarization and $\beta$-NMR beamline \cite{Kowalska2017,Gins2019}. The ion and laser beams enter from the left. The ions are represented by red circles with a plus sign. The neutral atoms are represented by blue circles. The polarization of the atom is represented by arrows. See text for further details.}
    \label{Fig:beamline}
\end{figure*}

$\beta$-NMR studies were performed on laser-polarized short-lived $^{26}$Na. The nuclei were produced at the ISOLDE facility at CERN \cite{Borge2018FocusPortrait}, in reactions induced by a 1.4-GeV proton beam of up to 2~$\mu$A, impinging every 3.6 s on a UC\textsubscript{x} (uranium carbide) target. After fast diffusion out of the heated target, sodium atoms were surface-ionized, accelerated to 50~keV and mass separated using the High Resolution Separator (HRS). The pure isotopic beam of $^{26}$Na, with an intensity of $2-5\times10^{7}$ ions/second, was transported to the laser polarization beamline \cite{Kowalska2017,Gins2019} shown in Fig. \ref{Fig:beamline}. 

\begin{figure}[btb!]
    \centering
        \includegraphics[width=0.49\textwidth]{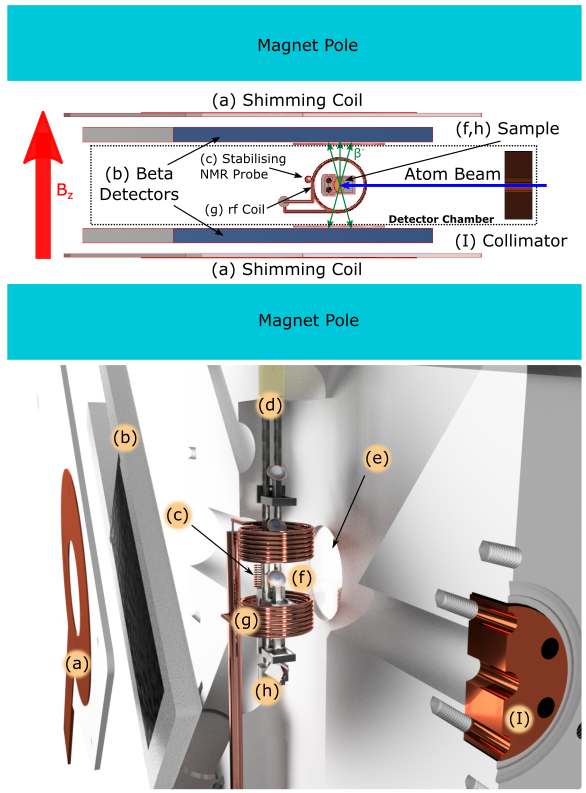}
        \caption{$\beta$-NMR detection chamber. Top: Schematic cross-section as viewed from the top. Bottom: 3D CAD exploded view. a) shimming coil to improve field homogeneity, b) $\beta$-particle detector (plastic scintillator) and Si photomultiplier, c) $^1$H-NMR probe to monitor and actively stabilize the magnetic field, d) sample ladder, e) $\beta$-particle window (100 $\mu$m aluminium), f) mica sample holder, g) main rf coil for NMR excitations, h) NaF crystal to optimise the degree of laser spin-polarization, i) 8 mm beam collimator. See text for further details.}
    \label{Fig:Chamber}
\end{figure}

There, the $^{26}$Na$^+$ beam is overlapped with circularly-polarized laser light. Next, it passes through a neutralization cell, where it picks up an electron as it travels through a vapour of stable $^{23}$Na. Over the next 1.5 m the neutral atomic $^{26}$Na beam is polarized via optical pumping in the D2 line at 589 nm~\cite{Kowalska2017}. This takes place in a weak guiding magnetic field of 2~mT (applied along the beam path), which defines the quantization axis and prevents the coupling of the electron spins to possible stray fields in the surrounding environment. Next, the atoms pass through a transitional field region of $\approx$ 10 - 20~mT, where the atomic spins undergo an adiabatic rotation towards the perpendicular magnetic field of the NMR magnet. The spin-polarized atoms pass through a collimator and reach a liquid sample located in a vacuum chamber that is placed between the poles of a Bruker BE25 electromagnet set to a field of 1.2~T  (Fig. \ref{Fig:Chamber}).  At this point, the nuclear and electronic spins are decoupled and the nuclear spin couples to the large static field. 

The liquid sample is deposited on a sample holder made of mica. The collimated atom beam and the holder have a diameter of 8~mm. Four such sample holders are attached to a sample ladder that can be moved in and out of the beam path. The emitted $\beta$ particles are registered in two pairs of thin organic scintillators, coupled to compact silicon photo-detectors. The sample at the center of the electromagnet is surrounded by a 30~mm diameter coil to which an rf signal can be applied. See Fig. \ref{Fig:Chamber} for details.

To record an NMR spectrum, such as the ones shown in Fig.~\ref{Fig:Freq_EMIM_BIM_plot}, 200 equally spaced rf frequencies are sequentially set. For each frequency, the $^{26}$Na beam is implanted over 200~ms following the proton-bunch impact. After the start of implantation the $\beta$ particles are counted for up to 1 s in the detectors at 0$^{\circ}$ and 180$^{\circ}$ to the direction of the magnetic field (left and right to the beam axis). From these counts the experimental $\beta$-decay asymmetry is determined, as a normalized difference in the counts, $(N_{0^{\circ}}-N_{180^{\circ}})/(N_{0^{\circ}}+N_{180^{\circ}})$. At the same time, the sample is irradiated with a continuous wave rf field of 0.03~mT and a frequency corresponding to the point in the scan. This procedure is repeated for consecutive proton bunches (arriving every 3.6 or 4.8~s), to allow most of the nuclei from the previous bunch to decay. If required by the signal-to-noise ratio, several spectra of the same sample can be recorded and summed.

To increase the precision of the NMR measurements to the ppm level, the magnetic field across the sample had to be homogeneous with a temporal stability at the ppm level during a measurement. To ensure the former, a weak magnetic field on the order of 0.02~mT was produced by two shimming coils placed in contact with the magnet poles \cite{Ginthard1960GENERATINGFELDS}. In this way the field homogeneity across the sample volume was improved by more than an order of magnitude in all three axes: 1 ppm along the symmetry axis of the magnet, 3 ppm in the vertical axis, and 5 ppm in the horizontal axis (ion-beam propagation). Since the magnetic field is symmetric with respect to the center of the sample, the remaining inhomogeneity contributes to a broadening of the resonance peak, without a significant shift in the resonance frequency, compared to a point-like sample. The temporal drift in the magnetic field was addressed using an active stabilization system based on the $^1$H resonance frequency measured in a tailor-made vacuum-compatible H$_2$O NMR probe. The 3-mm diameter probe was located just outside the main excitation rf coil, as shown in Fig.~\ref{Fig:Chamber}, with its middle only 25~mm away from the center of the sample. The resulting temporal stability was better than 1~ppm between sub-second and 24-h timescales, compared to drifts as big as 1 ppm/minute without it.

Previous $\beta$-NMR studies of the magnetic moments of short-lived nuclei have relied on solid-state hosts. For sodium, the studies were performed using a cubic NaF crystal which retained polarization for several dozen seconds, leading to NMR resonances with the width in the order of $10^{-3}$ of the resonance frequency \cite{Keim2000}. In comparison, with liquid-state hosts it is possible to obtain resonances with over two orders of magnitude smaller width (due to molecular tumbling within a liquid \cite{Abragam-nmr89}), whilst retaining the nuclear polarization long enough to employ $\beta$-NMR. 
Unfortunately, most liquid-state hosts used for NMR studies have a high vapour pressure, so when placed inside vacuum they either freeze or evaporate. However, room-temperature ionic liquids, which are salts in a liquid state at room temperature, have an extremely low vapour pressure \cite{Bier2010VapourLiquids}, which makes them suitable hosts for high-precision NMR studies in vacuum environments, as encountered in most $\beta$-NMR setups. For measuring the Larmor frequency of $^{26}$Na two different ionic liquids were selected: 1-ethyl-3-methylimidazolium dicyanamide (EMIM-DCA) and 1-butyl-3-methylimidazolium formate (BMIM-HCOO). The EMIM-DCA sample contained $\approx$ 1~$\mu$M of $^{23}$Na$^+$ while the BMIM-HCOO sample contained 0.5~M. Both samples were degassed slowly at $10^{-5}$ mbar pressure for several hours in a separate vacuum chamber. 20~$\mu$L of each solution was deposited as a 0.4 mm layer on one of the sample holders attached to the sample ladder. The ladder was then placed in the $\beta$-NMR chamber, as shown in Fig.~\ref{Fig:Chamber}, and the pressure inside was lowered slowly from atmospheric pressure to $10^{-5}$~mbar. The sample was oriented at 45 degrees to the atom beam. Due to the high viscosity of both liquids, the layer remained on each substrate at high vacuum for up to 24~h. 

\subsection{Conventional NMR on stable $^{23}$Na}

At the time of investigation, it was not possible to obtain a conventional NMR signal from $^{23}$Na at the $\beta$-NMR beamline. Therefore,  $^{23}$Na and $^1$H NMR spectra were recorded on a conventional NMR spectrometer. Our earlier systematic NMR studies showed that changing $^{23}$Na concentration from micro-molar to molar ranges and degassing for an extended period shifts the $^{23}$Na resonance by less than 0.5~ppm. This was taken as our experimental uncertainty for $^{23}$Na and degassing was not carried out during the measurements presented here. The field of 7.05~T was provided by a Bruker Avance DMX 300~MHz spectrometer and a basic pulsed-NMR scheme was applied (single $\frac{\pi}{2}$ rf pulse) on samples kept at room temperature. The sample preparation and Na$^+$ concentration were as close as possible to those in the $\beta$-NMR experiment: $\approx$ 1~$\mu$M in the EMIM-DCA sample and 0.5~M in the BMIM-HCOO sample. 

For the measurements, 200~$\mu$L of each solution were sealed inside a 3-mm diameter NMR tube. The tube was placed inside a 5-mm diameter tube filled with D$_2$O, whose $^2$H NMR signal was used to stabilise the magnetic field automatically during the measurements (field locking). $^1$H NMR resonances were also recorded within several minutes from $^{23}$Na spectra, using the same setup with two concentric tubes. Here, the 3-mm tube was filled with H$_2$O. Due to the way the field locking was performed, the magnetic field was the same for all measurements.

\section{Results}
In order to derive the nuclear magnetic moment from  the Larmor frequency in eqn. \ref{Eqn:larmor}, the effective magnetic field $B$ needs to be known. 
Since the external field $B_0$ is modified by the bulk magnetic susceptibility of the host and by the NMR shielding of the nucleus in the host $\sigma$, $B$ can be thus expressed as \cite{Becker-NMR2000} \begin{equation}
     B=B_0[1+(1/3-\alpha) \kappa)](1-\sigma),
    \label{Eqn:moment}
\end{equation}
where $\kappa$ is the volume magnetic susceptibility and $\alpha$ is the shape factor (see Appendix B for details).
It is advantageous to use an approach where $\alpha \kappa$ or/and $\sigma$ cancel out, so we started by determining the ratio $R$ of the magnetic moment of $^{26}$Na to that of $^{23}$Na in the same ionic liquid host
\begin{equation}
    R = \frac{\mu(^{26}\rm{Na})} {\mu(^{23}\rm{Na})} = \frac{\nu_L(^{26}\rm{Na}) I(^{26}\rm{Na})} {\nu_L(^{23}\rm{Na}) I(^{23}\rm{Na})} \frac{B(^{23}\rm{Na})} {B(^{26}\rm{Na})}.
    \label{Eqn:Ratio}
\end{equation}
This value is independent of the NMR shielding (which is the same for $^{26}$Na and $^{23}$Na), and includes only a correction due to the difference in bulk magnetic susceptibilities of our samples (Appendix B).

\begin{figure}[btb!]
	\begin{center}
		\includegraphics[width=80mm]{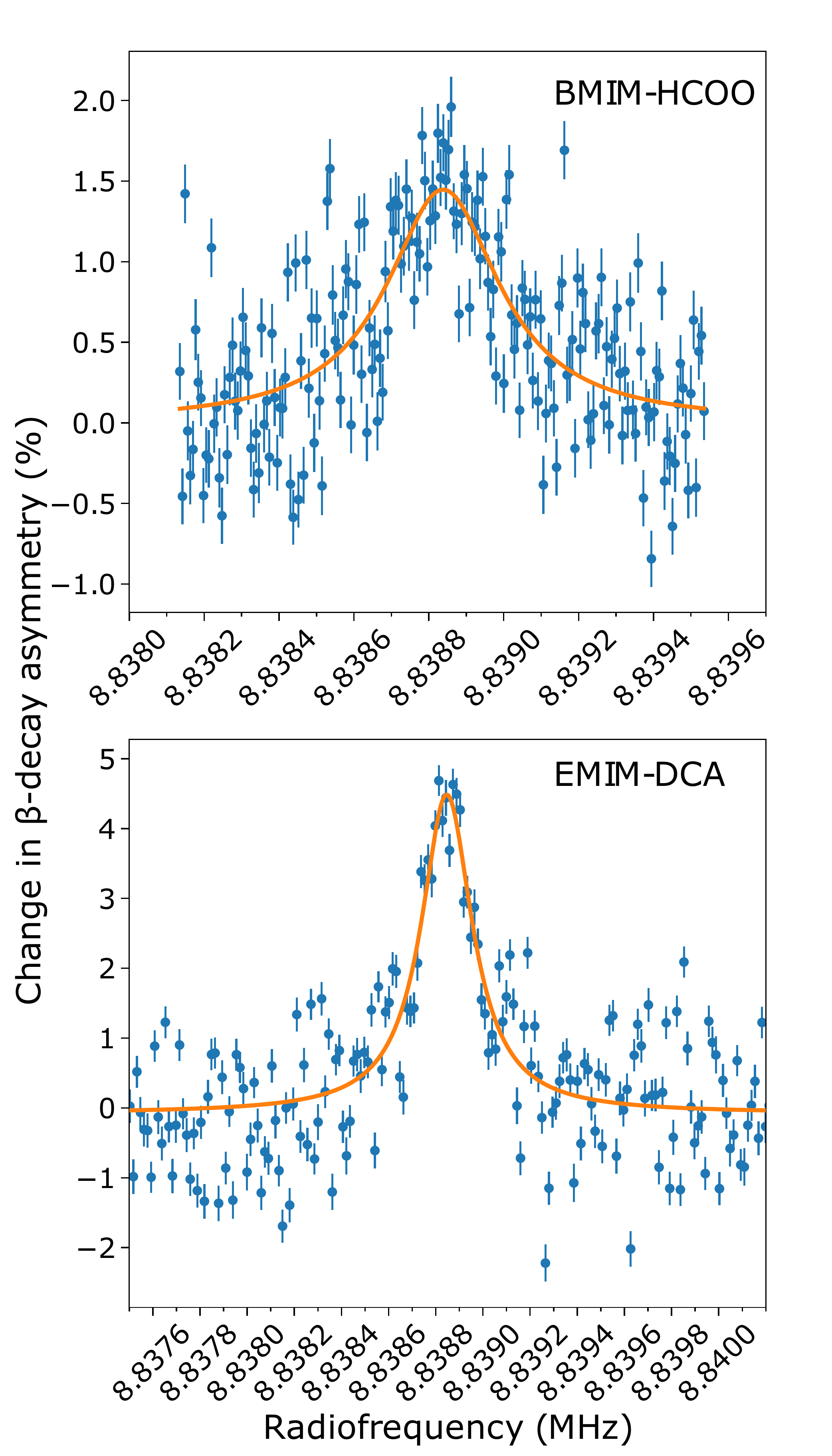}
		\caption{Examples of $^{26}$Na $\beta$-NMR spectra in BMIM-HCOO (top) and EMIM-DCA (bottom). 
		See text for the description of the fitting procedure. Note the different ranges of x and y axes. The fitted baseline (i.e. experimental asymmetry outside resonance) was subtracted for all data points for an easier comparison of the amplitudes of both signals. The magnetic field was locked to the same $^1$H frequency for both samples.}
		\label{Fig:Freq_EMIM_BIM_plot}
	\end{center}
\end{figure}

The $^{26}$Na $\beta$-NMR spectra in EMIM-DCA and BMIM-HCOO recorded at 1.2 T are shown in Fig. \ref{Fig:Freq_EMIM_BIM_plot}, while Tab.~\ref{Tab:Values} shows the corresponding Larmor frequencies, together with reference frequencies for $^{23}$Na at 7.05 T. For each measurement, several spectra were analysed, which differed in the observation time and in the coincident gate to determine the experimental $\beta$-decay asymmetry. The spectra were fitted with a flat baseline and Lorentzian profiles, which were expected due to a moderate rf power broadening. The data were also fitted using Gaussian profiles and a sloped baseline, with a negligible effect on the resonance frequency and its uncertainty.
 To extract the $^{26}$Na resonance frequencies shown in Tab. \ref{Tab:Values}, spectra with a 250 ms observation time were used, as they provided the highest signal-to-noise ratio and the smallest non-statistical scattering between data points. Such scattering is unavoidable in $\beta$-NMR measurements at facilities with a pulsed primary-beam structure (in our case, protons) in which the lifetime of the probe nuclei is comparable to the time between consecutive bunches of polarized atoms irradiated with a different rf field (e.g. $^{26}$Na). In such a case, $\beta$ particles emitted whose polarization has already relaxed in the sample contribute to a background which changes from one data point to another, due to variation in the time between consecutive proton pulses (alternating between 3.6 s and 4.8 s in our case). As a result, the normalized (reduced) sum of residuals $\chi^2_{red}$ was higher than 1, and the fitted frequency uncertainty was scaled by $\sqrt{\chi^2_{red}}$, following the procedure by the Particle Data Group \cite{Eidelman2004}. 

During the $^{26}$Na measurements, the $^{1}$H stabilising NMR probe had a resonance frequency of 52008500(30)~Hz. This was 1050(150)~Hz lower than when the probe was placed at the sample position in the middle of the magnet, which lead to a corrected frequency of 52009550(150)~Hz. During the $^{23}$Na measurements, the $^1$H NMR Larmor frequency was 300131415(100)~Hz. 

\begin{figure}[htb!]
	\begin{center}
		\includegraphics[width=85mm]{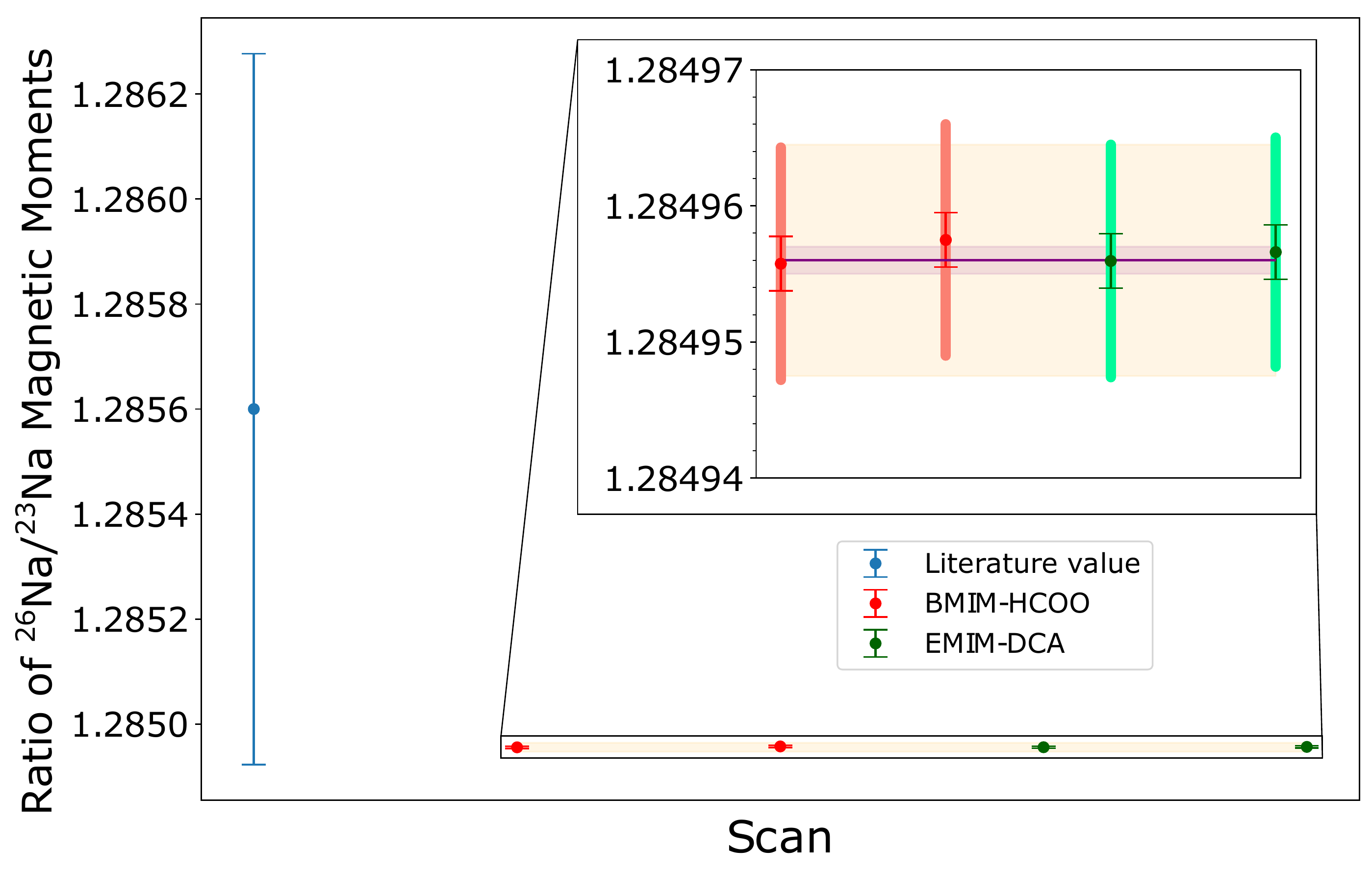}
		\caption{Ratio of the magnetic moments of $^{26}$Na and $^{23}$Na. Left: literature value \cite{huber78-Na-moments}, right:
		present study. Thin error-bar lines correspond to statistical uncertainty in the $^{26}$Na Larmor frequency and thick lines are the systematic uncertainties. The weighted average is represented by the purple line. The statistical uncertainty from all four measurements is indicated by the pink band, while the systematic uncertainty is shown by the broader orange band. For details, see text.}
		\label{Fig:Freq_ratio}
	\end{center}
\end{figure}

Using the above Larmor frequencies and the magnetic susceptibility correction from Appendix B, the derived value of $R$ for each measurement is shown in Tab.~\ref{Tab:Values}. The error in round brackets results from the statistical uncertainty on the $^{26}$Na resonance position. The systematic error present in all measurements is shown in square brackets, and includes systematic uncertainties in the resonance frequencies of $^1$H and $^{23}$Na, and the uncertainty of the magnetic susceptibility correction. Here, the biggest contribution by far is the error in the frequency of $^1$H during the $\beta$-NMR measurements, caused by the uncertainty in the position of the probe, which can be improved in the future. The final value of the ratio of $\mu(^{26}$Na) to $\mu(^{23}$Na) is $R = 1.284956(1)[8]$ or $R=1.284956(8)$ with the uncertainties combined. Fig.~\ref{Fig:Freq_ratio} shows the individual results in comparison to the literature value based on the hyperfine-structure measurement \cite{huber78-Na-moments}, which is two orders of magnitude less precise than our result. Our weighted average is indicated by the purple line. The purple shaded region represents the statistical uncertainty while the orange region represents the systematic uncertainties.

\begin{table}[]
	\centering
	\caption[Larmor frequencies of $^{26}$Na and $^{23}$Na]{Larmor frequencies of $^{26}$Na at 1.2 T and $^{23}$Na at 7.05 T in BMIM-HCOO and EMIM-DCA, and the resulting ratio of the magnetic moments, $R$, based on eqn. \ref{Eqn:Ratio}. Errors in round brackets are due to the statistical uncertainties in the resonance frequencies. For $R$, this includes only the uncertainty of the $^{26}$Na resonance frequency, while the square brackets are due to other contributions, including the uncertainty of the $^{23}$Na resonance frequency.}
	\renewcommand*{\arraystretch}{1.4}
	\begin{tabular}{cccc}
		\hline
		\hline
		Liquid host & $\nu_L(^{26}$Na)  (Hz) & $\nu_L(^{23}$Na) (Hz) & $R$     \\

		\hline
		BMIM-HCOO  & 8838826(14)  & \multirow{2}{*}{79390170(100)}  & 1.284956(2)[8] \\
		BMIM-HCOO   & 8838834(12) &   & 1.284957(2)[8] \\
		EMIM-DCA   & 8838838(10)  & \multirow{2}{*}{79390300(100)}  & 1.284955(2)[8] \\
		EMIM-DCA   & 8838847(13)  &   & 1.284957(2)[8] \\
		\hline
		\hline
	\end{tabular}
	\label{Tab:Values}
\end{table}

In order to determine $\mu$($^{26}$Na), a reliable reference $\mu$($^{23}$Na) value is needed.
In Nuclear Data Tables~\cite{Stone2005}, the values of $\mu$($^{26}$Na) based on Atomic Beam Magnetic Resonance (ABMR) and NMR experiments differ by 1.34$\times$10$^{-4}$ $\mu_N$,
which is much larger than the individual error bars. This introduces an uncertainty that is larger than that of the frequency-ratio measurement in our $\beta$-NMR experiment. 
The above discrepancy stems from applying an obsolete diamagnetic correction~\cite{NC_04} for the derivation of $\mu$($^{23}$Na) from the experiments. This inconsistency can be corrected using {\it ab initio} NMR shielding constants calculated for the species used in both experiments: a sodium atom in ABMR and an aqueous sodium ion in the NMR experiment. The technical details of NMR shielding calculations can be found in the Appendix A.

The NMR shielding in the sodium atom calculated using the Dirac-Hartree-Fock (DHF) method is 637.1 ppm.
The electron correlation contribution estimated using the Dirac-Kohn-Sham (DKS) method with various DFT functionals ranges from 0.06 ppm (PBE0) to 0.23 ppm (B3LYP). Coupled cluster codes for the NMR shielding of open-shell systems are not available. However, the accuracy of DKS correlation contributions can be estimated by the NMR shielding in the closed-shell Na$^+$ ion and the difference between the electron correlation contributions in the sodium atom and sodium ion from the literature~\cite{Pyper_1985}. The non-relativistic CCSD(T) correlation contribution calculated for the sodium ion, $-0.08$ ppm, should not differ from the NMR shielding in the sodium atom by more than 0.09 ppm~\cite{Pyper_1985}. 
All presented correlation contributions suggest that the electron correlation effects for the NMR shielding in the sodium atom are small. Therefore the NMR shielding in the sodium atom can be approximated with a very good accuracy by the DHF value.
The electron correlation contributions can be used as an error estimate. Our final NMR shielding in the sodium atom, 637.1(2) ppm, is consistent with the shielding in ref.~\cite{mason1987}. 
The NMR shielding for the solvated sodium ion was approximated by the NMR shielding in a six-coordinated Na$^+$(H$_2$O)$_6$ complex (the prevalent coordination number according the experiment~\cite{Galib2017}). 
The five coordinated Na$^+$(H$_2$O)$_5$ complex was used to estimate the error of the NMR shielding due to the structural uncertainty.
The NMR shielding constants calculated using non-relativistic and relativistic approximations for the Na$^+$(H$_2$O)$_6$ complex are shown in Tab.~\ref{sigmatab}. The Hartree-Fock and DHF NMR shielding, electron correlation contributions 
($\Delta^{\mathrm{CCSD}}_{corr}$, $\Delta^{\mathrm{CCSD(T)}}_{corr}$) and relativistic contributions 
($\Delta^{\mathrm{DHF}}_{rel}$, $\Delta^{\mathrm{DKS}}_{rel}$) reach good convergence with the basis size.
The final NMR shielding for the Na$^+$(H$_2$O)$_6$ complex, 582.0 ppm, is a composite value of (i) the non-relativistic shielding calculated using the CCSD(T) method, (ii) relativistic correction ($\Delta^{\mathrm{DKS}}_{rel}$), and (iii) the PCM solvent contribution ($\Delta^{\mathrm{PCM}}_{solv}$). All contributions entering the final NMR shielding were calculated using the quadruple-$\zeta$ (QZ) basis set.

The systematic error of the NMR shielding in the Na$^+$(H$_2$O)$_6$ complex was evaluated as the square root of the sum of squares of the following errors. The structural uncertainty (2 ppm) was evaluated as the difference between the CCSD NMR shielding for aqueous sodium complexes with the coordination number of five and six.
The basis set incompleteness error (1 ppm) was estimated from the variations of the NMR shielding constants calculated using non-relativistic HF method with Dunning and Jensen basis set series.
The coupled cluster expansion truncation error was approximated by $\Delta^{\mathrm{CCSD(T)}}_{corr} \approx 1$ ppm. 
Considering the convergence of the PCM solvent contribution ($\Delta^{PCM}_{solv}$), the error was estimated to be 1 ppm. The systematic error introduced by assuming an additivity of the
electron correlation and the relativistic effects is negligible as indicated by the small difference between the $\Delta^{\mathrm{DHF}}_{rel}$ and $\Delta^{\mathrm{DKS}}_{rel}$ relativistic corrections.

The final approximation of the NMR shielding of the aqueous sodium ion is {(582.0 $\pm$ 2.6) ppm.} This result is consistent with the NMR shielding in ref.~\cite{aadkakkmjcpl532}, but in the present study, the error bar was reduced by a factor of four. This was achieved by calculations with much larger basis sets, which led to a better convergence of all contributions.

Table~\ref{mmref} presents the new values of the $^{23}$Na reference magnetic moment re-derived using our new NMR shielding constants. The ABMR-based magnetic moment was obtained using our {\it ab initio} NMR shielding of the sodium atom and the original ABMR experiment~\cite{Beckmann1974}. The NMR-based magnetic moment was re-derived using (i) our {\it ab initio} NMR shielding of the aqueous sodium ion, (ii) the experimental frequency ratio 0.26451900~\cite{rkhedbsmcmrgpgmric40} of $^{23}$Na in 0.1 M NaCl water solution to the proton in tetramethylsilane (TMS), (iii) the reference proton magnetic moment $\mu(^1\rm{H})$ = 2.792847348(7) $\mu_N$~\cite{Mooser2014}, and (iv) the reference NMR shielding of the proton in TMS $\sigma(^1\rm{H})$ = 33.480 $\pm$ 0.5 ppm~\cite{Garbacz2012}.

The newly extracted ABMR- and NMR-based values of $^{23}$Na nuclear magnetic dipole moment are now consistent within the error bars and the discrepancy between them was decreased by a factor of $\approx$30.

For the derivation of the $^{26}$Na nuclear magnetic dipole moment, the NMR-based $^{23}$Na nuclear magnetic dipole moment was used, because the corresponding NMR shielding calculations for aqueous sodium complexes are based on a better approximation and the error bar was estimated more rigorously. The resulting $^{26}$Na nuclear magnetic dipole moment is 2.849390(20) $\mu_N$ (Tab.~\ref{Tab:Moments}).

The new $^{26}$Na nuclear magnetic dipole moment is consistent with the previous experimental value based on the hyperfine-structure measurement \cite{huber78-Na-moments} within the error bar, but the present experiment and \textit{ab initio} calculations improved its accuracy by two orders of magnitude, to 7 ppm. The largest contribution to this error bar comes from the uncertainty in the position of the $^1$H NMR probe during the $\beta$-NMR experiment, which is 2 times larger than the uncertainty from NMR shielding and 3 times larger than the other experimental uncertainties. Experimental upgrades to provide a rigorous determination of the probe position could reduce the uncertainty of the $^{26}$Na magnetic moment to the level of accuracy reached for the stable $^{23}$Na.

\begin{table}
\caption{Sodium NMR shielding in the Na$^+$(H$_2$O)$_6$ complex.}
\begin{ruledtabular}
\begin{tabular}{lrrrr}
                                &   DZ$^a$                &   TZ$^a$         &   QZ$^a$       \\
\hline                          
 HF                             &   578.588           &  578.814     &  579.150   \\
CCSD                               &   571.625           &  573.837     &  574.140   \\
CCSD(T)                            &   571.011           &  572.909     &   573.127         \\
$\Delta^{\mathrm{CCSD}}_{corr}$    &   -6.963            &  -4.977      &  -5.010    \\
$\Delta^{\mathrm{CCSD(T)}}_{corr}$ &   -0.614             & -0.928      &  -1.013          \\
PBE0+PCM                           &   563.355          &  565.609    &  568.197 \\
PBE0                            &   564.406          &  565.472    &  567.533  \\
$\Delta^{\mathrm{PCM}}_{solv}$   &   -1.051           &  0.137      &  0.664    \\
\hline                          
DHF                        &   586.860           &  587.263     &  587.346   \\
DHF$^{b}$                     &   578.980           &  579.089     &  579.151   \\
$\Delta^{\mathrm{DHF}}_{rel}$            &   7.880             &  8.174       &  8.195     \\
DKS/PBE0                  &   574.842           &  574.822     &  574.848   \\
DKS/PBE0$^{b}$                &   567.007           &  566.694     &  566.688   \\
$\Delta^{\mathrm{DKS}}_{rel}$   &   7.835             &  8.128       &  8.160     \\
\end{tabular}
\begin{flushleft} $^a$ for non-relativistic calculations cc-pCVXZ.cc-pVXZ basis set series (X = D, T, Q) are used; for relativistic DHF and DKS calculations uncontracted ucc-pCVXZ.ucc-pVXZ basis set series are used\\
$^b$ non-relativistic limit obtained with the speed of light re-scaled by factor of 20
\end{flushleft} 
\label{sigmatab}
\end{ruledtabular}
\end{table}

\begin{table}
\caption{$\mu$($^{23}$Na)/$\mu_N$ reference nuclear magnetic dipole moment from ABMR and NMR experiments} 
\begin{ruledtabular}
\begin{tabular}{lllll}
                  &  old reference~\cite{Stone2005}  &   This work   \\
\hline
ABMR              &  +2.217522(2)           &   2.217495(2)$^a$    \\
NMR               &  +2.2176556(6)          &   2.217500(7)$^b$    \\
\end{tabular}
$^a$ using the original ABMR experiment~\cite{Beckmann1974} and NMR shielding of the sodium atom (637.1 $\pm$ 0.2) ppm \newline
$^b$ using the standard NMR frequency ratio of $^{23}$Na in NaCl water solution to proton in TMS~\cite{rkhedbsmcmrgpgmric40}
and NMR shielding of Na$^+$(H$_2$O)$_6$ {(582.0 $\pm$ 2.6) ppm.} See the text for details on NMR shielding calculations.
\label{mmref}
\end{ruledtabular}
\end{table}

\begin{table}[t]
        \centering
        \caption{Magnetic moments of $^{23,26-31}$Na determined in this work, compared to literature values \cite{huber78-Na-moments,Keim2000}, and other nuclear properties relevant for NMR.}
        \label{Tab:Moments}
        \renewcommand*{\arraystretch}{1.4}
        \begin{tabular}{ccrrcl}
                \hline
                \hline
                ~~Isotope~ &~$I$~   &~$t_{1/2}$(ms)~& $Q (mb)$ & old $\mu$ $(\mu_N)$ ~& new $\mu$ ($\mu_N$)~ \\
                \hline
                $^{23}$Na &     3/2 &   stable      &     +106(1)    &            & 2.217500(7)$^a$ \\
                $^{26}$Na &     3   &   1071        &    -5.3(2)    &   2.851(2)  & 2.849390(20)$^b$\\
                $^{27}$Na &     5/2 &   301         &    -7.2(3)    &   3.894(3)  & 3.89212(24)    \\
                $^{28}$Na &     1   &   31          &   +39(1)      &   2.420(2)  & 2.41844(19) \\
                $^{29}$Na &     3/2 &   44          &   +86(3)      &   2.457(2)  & 2.45535(17) \\
                $^{30}$Na &     2   &   48          &               &   2.069(2)  & 2.0681(11) \\
                $^{31}$Na &     3/2 &   17          &               &   2.298(2)  & 2.29670(17) \\
                \hline
                \hline
        \end{tabular}
\\      
$^a$ corrected $\mu(^{23}$Na) based on NMR experiment, Table \ref{mmref}\\
$^b$ based on our improved ratio of $\mu(^{26}$Na)/ $\mu(^{23}$Na)\\
\end{table}

Magnetic moments which have been linked to $^{26}$Na can also benefit from the improved accuracy of $\mu$($^{26}$Na). This is the case for $^{27-31}$Na, which were investigated using $\beta$-NMR in solid-state hosts at the collinear laser spectroscopy beamline at ISOLDE \cite{Keim2000}, and whose $g$-factors $g_I=\mu/(I\mu_N)=\gamma \hbar / \mu_N$ 
were referenced to that of $^{26}$Na. Table ~\ref{Tab:Moments} presents our new values of the $^{23}$Na and $^{26}$Na magnetic moments, as well as the $^{27-31}$Na magnetic moments obtained using our improved $\mu(^{26}$Na) and the aforementioned $g$-factors. Literature magnetic moments~\cite{huber78-Na-moments,Keim2000} are also shown for comparison.

The new values of the $^{27-29,31}$Na magnetic moments have a relative uncertainty of 70~ppm. This is a ten-fold improvement compared to the values deduced in \cite{Keim2000} and up to 50 times more accurate than the values tabulated in the latest compilation of nuclear magnetic dipole and electric quadrupole moments~\cite{Stone2005} (for $^{30}$Na, it is respectively two \cite{Keim2000} and 10 \cite{Stone2005} times smaller). Previously, the uncertainty for $^{27-31}$Na was dominated by the precision in the magnetic moment of the reference $^{26}$Na. At present, it is determined by the uncertainty in the $^{27-31}$Na $\beta$-NMR resonance frequency in solid-state hosts. If new measurements in liquid hosts are performed, this contribution could be decreased further to the ppm level.

\section{Discussion and future perspectives}

To determine precise and accurate, shielding-corrected magnetic moments, two independent steps are needed. First, the Larmor frequency of the radioactive probe is measured relative to that of a stable NMR probe, e.g. $^1$H or $^2$H in water. This procedure removes the need for reference measurements relative to another radioactive probe nucleus, which is the current (time consuming) reference method used in $\beta$-NMR. Furthermore, by using an ionic liquid as the host for the radioactive probe, a very precise Larmor frequency can be obtained, from which a precise (but still uncorrected) magnetic moment of a short-lived nucleus can be deduced relative to that of the stable ($^1$H or $^2$H) probe. To correct for the NMR shielding in the host, two procedures are possible. The NMR shielding in the host material can be calculated using modern calculation methods (if possible), or alternatively an independent NMR measurement has to be performed for the stable isotope of the element in the same host, again relative to the H-reference.  The latter approach was used here.  The final accuracy on the magnetic moment will then depend on the accuracy of the moment of the stable isotope, which can be deduced from former high-precision measurements in atoms, molecules and liquids, in combination with state-of-the-art shielding calculations (as performed here).

The accurate magnetic moments of $^{26-31}$Na presented above, together with that of $^{23}$Na, provide a set of NMR probes connected through the same NMR shielding. In this way conventional NMR and the ultra-sensitive $\beta$-NMR can be used to provide complementary information on chemical and biological processes, by probing different  timescales and different nucleus-environment interactions (see Tab. \ref{Tab:Moments}).
For example, with the very short-lived $^{28}$Na one can probe processes with ms timescales, with longer-lived $^{26}$Na -- timescales of seconds, while stable $^{23}$Na has a much longer observation window. Furthermore, quadrupole moments of $^{26}$Na and $^{27}$Na are respectively 20 and 15 times smaller compared to the stable $^{23}$Na. This results in a weaker interaction with the gradient of the electric field \cite{Slichter1990ElectricEffects}, leading to longer relaxation times and narrower resonances. This should permit the observation of NMR signals in hosts which display broad $^{23}$Na resonances due to a fast quadrupolar relaxation.

The approach presented here can be directly applied to other isotopic chains, thus expanding the palette of nuclei available for NMR spectroscopy. It can be combined with several techniques to polarize spins of short-lived nuclei. Some elements are easily polarized using element-specific laser optical pumping, as proven for several alkali and alkali-earth elements \cite{NeugartNeyens2006-moments}. At the same time, universal polarization methods, such as pickup of polarized thermal neutrons, projectile-fragmentation or low-energy nuclear reactions can be also used to produce polarized samples of radioactive isotopes, see \cite{Ackermann1985,Nojiri1985,Neyens2003NuclearNuclei} and references therein.

Accurate magnetic moments of $\beta$-NMR probe nuclei are setting foundations for a novel referencing scheme in $\beta$-NMR spectroscopy. The method is based on measuring two Larmor frequencies simultaneously: for the radioactive probe in the chosen host material and a stable NMR probe like $^1$H or $^2$H in a water placed in the experimental setup near the probe of interest. In this scheme, the absolute NMR shielding $\sigma_{X}$ instead of a chemical shift could be measured directly, following eqns. \ref{Eqn:larmor} and \ref{Eqn:moment}: 
\begin{equation}
\sigma_{X} = 1 - \frac{\nu_{X}}{\nu_{Y}}\frac{|\mu_Y|}{|\mu_X|}\frac{I_X}{I_Y} \frac{1+(1/3 - \alpha_Y) \kappa_Y} {1+ (1/3 - \alpha_X) \kappa_X} (1-\sigma_{Y}) .
\label{eqn:referencing}
\end{equation}
where $X$ is the $\beta$-NMR probe nucleus and $Y$ is the reference conventional nucleus (e.g $^1$H in water). A description of the correction due to the difference in the bulk magnetic susceptibilities $\alpha \kappa$ is presented in Appendix B.
Here, the $\beta$-NMR probe nucleus is related to the conventional NMR reference nucleus, which establishes a bridge between $\beta$-NMR spectroscopy and conventional NMR spectroscopy. This scheme offers the possibility to reference radioactive nuclei shielding to the stable nuclei not only within the isotopic chain, but also between different elements. This removes the dependence of $\beta$-NMR spectroscopy on the ambiguous and often {\it ad hoc} standards defined for every element separately \cite{Jackowski2010b}.

In this novel referencing scheme, the uncertainty of the NMR shielding $\sigma_X$ of $\beta$-NMR nuclei in different hosts, derived from eqn. \ref{eqn:referencing}, will be defined primarily by the uncertainty in their magnetic moment. 
For $^{26}$Na, using the old value of $^{26}$Na magnetic moment leads to $^{26}$Na NMR shielding values with a $\pm$700 ppm error bar, which is about 10 times larger than the full range of chemical shifts for sodium \cite{Hanusa2015}. In comparison, our new magnetic moment of $^{26}$Na will lead to a 100 times more accurate shielding values ($\pm$7 ppm), which will be sufficient to distinguish between different sodium binding sites, see e.g. \cite{Schmeisser-gutman-donors-23NaNMR-IL-2012}, and will enable comparisons to theoretical Na NMR shielding values \cite{tossell-Na-shielding-jpcb2001, MCMILLEN-Na-shielding2010,wong2009}. 

All of the above innovations institute novel applications for $\beta$-NMR in chemistry and biology. One such application is the interaction of metal ions with biomolecules \cite{Crichton2019BioInorgChem,Crichton2020PractBioInorgChem}, which is important for the functions of living organisms (especially metal-ion mediated folding of proteins \cite{Sigel2001-proteins} and nucleic acids \cite{Sigel2012-metalsDNA}). For example, half of the proteins in our body contain metal ions, but their interactions and factors influencing them are still not fully understood. This is because many metal ions are silent for most spectroscopic techniques \cite{Crichton2020PractBioInorgChem} and are very challenging for conventional NMR \cite{mason1987,Riddell2016}. Yet, in NMR, metal nuclei are often very sensitive to small changes in geometry and coordination number, which gives rise to dozen-ppm shifts in resonance frequencies for many metals \cite{Riddell2016,Hanusa2015}. The application of $\beta$-NMR will allow this field to profit from up to a billion times increased sensitivity and  access to readily available $\beta$-NMR probe nuclei with smaller or even no quadrupolar moment (see e.g.  \cite{Keim2000,Neyens2005MeasurementState}), giving rise to longer relaxation times and narrower resonances.

Using the advances presented here, pilot applications in biology are already planned. Among the biologically relevant metal ions, sodium and potassium play an important role in the formation and dynamics of special DNA structures, G-quadruplexes, which are promising targets for anti-cancer therapies \cite{Carvalho2020}. Our present work has prepared $^{26}$Na to be an immediately applicable $\beta$-NMR probe to address this topic \cite{Karg2020proposal,Wu2003-JACS-K-NMR-GQ,WONG2005-GQ,wong2009}. Presently, we are also exploring the most suitable potassium probes for G-quadruplex studies \cite{Karg2020proposal} and isotopes of several other elements relevant to protein folding \cite{Jancso2017}.

In a very different field, namely in nuclear structure, our research paves the way for addressing the open question about the distribution of neutrons inside atomic nuclei \cite{Hagen2016,Thiel_neutron-skin-2019}. The neutron distribution impacts the properties of neutron stars \cite{aumann-n-skin-EOS-2017}, determines the limits of the nuclear landscape \cite{Erler2012-neutron-drip-nature}, and is responsible for novel phenomena and exotic structures in unstable nuclei \cite{tanihata1985-PRL-11li}. It is especially important for light neutron-rich `halo' nuclei, consisting of a compact nuclear core and one or several loosely bound `halo' neutrons which are spatially extended \cite{Tanihata_1996,Al-Khalili2004}. As neutrons do not carry an electric charge, compared to protons their distribution is much more difficult to be determined experimentally. However, because the neutron distribution is closely related to the distribution of nuclear magnetism, it can be addressed via the hyperfine anomaly, by combining the accurate magnetic moment with an accurate hyperfine structure measurement \cite{Stroke-HypInt-BW2000}. 
For example, in $^{11}$Be the magnetism is mostly due to the `halo' neutron \cite{Takamine2016,Puchalski2014}, so the hyperfine anomaly provides a direct probe of the halo structure \cite{Parfenova2005, Puchalski2014}. Because the hyperfine structure of $^{11}$Be is already known with high accuracy \cite{Takamine2016}, the only missing experimental input to derive the neutron distribution from the hyperfine anomaly is an accurate value of the magnetic moment of $^{11}$Be, which can be achieved by applying the procedure presented in this work.

\section{Conclusions} 

In summary, using $^{26}$Na as an example, we have presented the first determination of a magnetic moment of a short-lived nucleus with ppm accuracy. This represents an improvement by two orders of magnitude in comparison with a previous experiment and other $\beta$-NMR based measurements of magnetic moments. The procedure described in this article represents a general protocol for measurements of magnetic dipole moments of polarized $\beta$-decaying nuclei with high accuracy, reaching the accuracy for stable nuclei.

The innovations presented here brings the following advances for the ultra-sensitive $\beta$-NMR technique: 
(i) Elimination of the dependence of $\beta$-NMR spectroscopy on ambiguous and often {\it ad hoc} references. As a result, the uncertainty related to the $\beta$-NMR reference measurement can be removed from the analysis. In addition, the direct comparison of $\beta$-NMR and conventional NMR data bridges these two techniques.
(ii) Saving scarce resources of radioactive beam for acquisition of more $\beta$-NMR data on the samples of interest, since a reference measurement on a $\beta$-NMR probe is not required. This will accelerate the application of $\beta$-NMR spectroscopy as an analytical tool.
(iii) Link to {\it ab initio} predictions through the direct measurement of NMR shielding for $\beta$-NMR probes. This will facilitate the interpretation of $\beta$-NMR experiments.

These novel features have the potential to transform $\beta$-NMR spectroscopy into a more widely applicable technique, based on a palette of ultra-sensitive $\beta$-NMR probes with accurate magnetic moments, allowing to address problems that range from neutron distribution in exotic nuclei to interactions of metal ions with biomolecules.

\section{Acknowledgements}
This work was supported by the European Research Council (Starting Grant 640465), the UK Science and Technology Facilities Council (ST/P004423/1), FWO-Vlaanderen in Belgium (G0B3415N), KU Leuven (GOA 15/010), EU project ENSAR2 (654002), Slovak Research and Development Agency grant (APVV-15-0105), European Regional Development Fund, Research and Innovation Operational Programme (ITMS2014+: 313011W085), Polish National Science Centre OPUS research grant (2017/27/B/ST4/00485), the Ministry of Education of Czech Republic (LM2015058), the Wolfgang Gentner Programme of the German Federal Ministry of Education and Research (05E15CHA), and the Swiss Excellence Scholarship programme. Computational resources of the Slovak Academy of Sciences and the Slovak University of Technology were used (projects ITMS 26230120002 and ITMS 26210120002).
We thank the assistance of the ISOLDE technical team and that of L. Hemmingsen from Copenhagen University, M. Walczak from Poznan University of Technology, K. Szutkowski from A. Mickiewicz University in Poznan, M. Jankowski, R. Engel, and W. Neu from Oldenburg University, H. Heylen, A. Beaumont, M. Van Stenis from CERN, V. Araujo from KU Leuven, A. Zhuravlova from Kiev University, M. Piersa and E. Adamska from Warsaw University, J. Klimo, R. Urban, S. Komorovsky, G. Kantay, J. Kranjak from the Slovak Academy of Sciences, E. Sistare from Geneva University, and M. Jaszu{\'n}ski from the Polish Academy of Sciences.

\section{Appendix A - {\it Ab initio} NMR shielding calculations}

NMR shielding in the sodium atom with the doublet electronic ground state was calculated using the Dirac-Hartree-Fock (DHF) method applying the paramagnetic NMR theory for open-shell systems~\cite{Malli2009,Pyper_1985,Komorovsky2013pnmr}. 
Dyall-VXZ~\cite{Dyall2006} basis set series were used (X = D, T, Q represents double-$\zeta$, triple-$\zeta$ and quadruple-$\zeta$ basis sets).\\
According to a recent experiment~\cite{Galib2017}, the coordination number of the aqueous Na$^+$ ion depends on the NaCl solution concentration and varies between 5 and 6. Therefore, NMR shielding of the Na$^+$ ion in the aqueous solution was calculated for model Na$^+$(H$_2$O)$_5$ and Na$^+$(H$_2$O)$_6$ complexes. Their structures were optimized using Density Functional Theory (DFT) with the B3LYP density functional~\cite{dft:becke88,dft:lyp1,dft:vwn} and Def2-TZVP basis set~\cite{weigend2005a}. The D3 dispersion correction~\cite{Grimme2011} was applied. 
A distorted octahedral structure (D$_{2h}$ symmetry) was obtained for the Na$^+$(H$_2$O)$_6$ complex, with an average Na-O distance of 2.386~\AA. For Na$^+$(H$_2$O)$_5$ the corresponding structure was found to be a trigonal bipyramid (C$_{2v}$ symmetry) with an average Na-O distance of 2.368~\AA. 
The average Na-O distances for both structures are in good agreement with the experimental Na-O distances obtained with two different experimental methods giving 2.384 $\pm$ 0.003~{\AA} and 2.37 $\pm$ 0.024~{\AA}~\cite{Galib2017}.

NMR shielding constants for aqueous sodium complexes were calculated using the non-relativistic coupled cluster (CC) method with single and double excitations (CCSD) and with non-iterative triple excitations CCSD(T)~\cite{Gauss1995,Watts1993}. All electrons were correlated. Dunning core-valence basis set series cc-pCVXZ~\cite{schuchardt2007a} were used for sodium and valence series cc-pVXZ~\cite{thdjcp90} for hydrogen and oxygen, combining basis sets with the same cardinal number X (X = D, T, Q). 
In order to estimate the error due to incompleteness of the basis set, the pcS-n basis set series by Jensen~\cite{Jensen2008a} was also used.
In all NMR shielding calculations Gauge-Including Atomic Orbitals (GIAO)~\cite{kwjfhppjacs112} were used.

The effect of the water solvent (outside the first solvation shell) on the NMR shielding in the sodium complex was incorporated by the polarized continuum model (PCM) COSMO~\cite{P29930000799}. This effect was evaluated using DFT with the PBE0 functional~\cite{Perdew1996,Adamo1999}. 
The water dielectric constant of 78 was used in this implicit solvent model.

Relativistic corrections were calculated as the difference between the relativistic NMR shielding and the corresponding non-relativistic limit using two different methods: the DKS method with the PBE0 functional and the DHF method. The non-relativistic limit was obtained by re-scaling the speed of light in the Hamiltonian by a factor of 20. In the relativistic calculations, the Dunning basis sets were fully uncontracted and a restricted magnetic balance scheme was employed to generate the small component basis set ~\cite{Komorovsk2008,Komorovsk2010}. The nucleus was modeled by a Gaussian charge distribution~\cite{visscher1997}. 

For the structure optimization and for non-relativistic DFT calculations of NMR shielding constants the NWChem package was used~\cite{nwchem}. Non-relativistic coupled cluster NMR shielding calculations were carried out in the CFOUR~\cite{cfour:13} package. For relativistic NMR shielding calculations, the ReSpect~\cite{ReSpect-5.0.1,Repisky2020} program was used.

\section{Appendix B - Effect of bulk magnetic susceptibility}

When using NMR to determine accurate nuclear magnetic moments or absolute NMR shielding, one should consider the differences in bulk magnetic susceptibility between the samples \cite{Becker-NMR2000} (see. eqn. \ref{Eqn:moment} and \ref{eqn:referencing}). This effect depends on the volume magnetic susceptibility of the host material $\kappa$ and on the geometry of the sample, reflected in the shape factor $\alpha$. For the shapes used in our studies, $\alpha \approx 0$ for the $^{23}$Na and $^1$H samples in conventional NMR, and for $^{26}$Na in $\beta$-NMR at CERN (cylinders parallel to the magnetic field \cite{Becker-NMR2000} and a disc perpendicular to the magnetic field \cite{schenck-MedPhys-magn-susc-1996}, respectively), whereas $\alpha\approx 1/2$ for the $^1$H probe used at CERN (cylinder perpendicular to the field \cite{Becker-NMR2000}. 

In eqn. \ref{Eqn:Ratio} for the ratio $R$ of the magnetic moments, the magnetic susceptibility corrections for $^{23}$Na and $^{26}$Na cancel out in the term $\nu_L(^{26}\rm{Na}) / \nu_L(^{23}\rm{Na})$, due to the same $\alpha$ and $\kappa$. At the same time, $B(^{23}\rm{Na}) / B(^{26}\rm{Na}) =  \nu_L(^1H) / \nu_L^{'}(^1H) \times (1+ \Delta)$, with 
$\Delta = (1- \frac{1}{6}\kappa_{H_2O}) / (1+ \frac{1}{3} \kappa_{H_2O}) - 1$, and prime denoting the measurement at CERN. Using $\kappa_{H_2O}=-9.04\times 10^{-6}$, $\Delta \approx +4.5 \pm 0.5$ ppm, where we assumed a 10 \% uncertainty in the shape factors due to the finite size of the samples.

When using eqn. \ref{eqn:referencing} to measure NMR shieldings with $\beta$-NMR, one must consider the shape factor $\alpha$ and the volume susceptibility $\kappa$ for the host of the reference nucleus (water in our case) and that of the $\beta$-NMR nucleus.

\bibliographystyle{PRX}

\end{document}